\documentclass[english,aps,prb,amsmath,twocolumn,preprintnumber,superscriptaddress,showpacs,notitlepage,nofootinbib,amssymb,floatfix]{revtex4-2}
\usepackage[T1]{fontenc}
\usepackage[latin9]{inputenc}
\usepackage{mathptmx,newtxtext,newtxmath,xspace,amsfonts}
\usepackage{amsbsy,bm,bbold,amsmath}
\usepackage{graphicx,color,xcolor,epsfig,rotate}
\usepackage{fancyhdr}
\usepackage[colorlinks=true, 
            linkcolor=blue, 
            urlcolor=blue,
            citecolor=blue]{hyperref}
\usepackage[capitalise]{cleveref}

\usepackage{amssymb}
\usepackage[colorlinks=true, 
            linkcolor=blue, 
            urlcolor=blue,
            citecolor=blue]{hyperref}
\usepackage{soul}
\usepackage{cancel}
\usepackage{babel}
\usepackage{appendix}
\usepackage{pifont,textcomp,enumerate,threeparttable,setspace,subfigure}

\makeatother

\begin{document}
\title{Probing Chiral Kitaev Spin Liquids via  Dangling Boundary Fermions}
\author{Shang-Shun~Zhang}
\affiliation{Department of Physics and Astronomy, The University of Tennessee,
Knoxville, Tennessee 37996, USA}
\author{G\'abor B. Hal\'asz}
\affiliation{Materials Science and Technology Divison, Oak Ridge National Laboratory, Oak Ridge, Tennessee 37831, USA}
\affiliation{Quantum Science Center, Oak Ridge, Tennessee 37831, USA}
\author{Cristian~D.~Batista}
\affiliation{Department of Physics and Astronomy, The University of Tennessee,
Knoxville, Tennessee 37996, USA}
\affiliation{Neutron Scattering Division and Shull-Wollan Center, Oak Ridge
National Laboratory, Oak Ridge, Tennessee 37831, USA}

\date{July 2024}

\begin{abstract}
Identifying experimental probes capable of diagnosing extreme quantum behavior is widely regarded as one of the foremost challenges in modern condensed matter physics. Here, we propose a novel approach for detecting chiral Kitaev spin liquid states through measurements of the local dynamical spin structure factor on the boundary using scanning tunneling microscopy (STM). We specifically focus on unpaired (``dangling'') Majorana fermions, which naturally emerge along boundaries of Kitaev spin liquids, and can serve as indicators of chiral boundary modes under broad conditions, thereby offering a clear signature of these exotic quantum states.
\end{abstract}


\maketitle

\section{Introduction}

The exact solution of the famous honeycomb Kitaev model \cite{Kitaev06} has sparked significant efforts to discover materials that host chiral Kitaev spin liquids (KSLs). These spin liquids are characterized by topologically robust chiral edge modes propagating along the boundary, and may support non-Abelian bulk excitations binding Majorana zero modes. Notably, it has been shown that the bond-directional spin interactions of the Kitaev model naturally arise between effective spin-1/2 magnetic moments in spin-orbit-coupled $4d$ and $5d$ materials~\cite{jackeli2009mott}. In turn, this insight has driven the discovery of several candidate materials where the microscopic spin Hamiltonian is believed to closely approximate the Kitaev model~\cite{rau2016spin, trebst2017kitaev, hermanns2018physics, takagi2019concept}. 

These ``Kitaev materials'' include the honeycomb iridates, such as Na$_2$IrO$_3$~\cite{singh2010antiferromagnetic, liu2011longrange, choi2012spin, ye2012direct, comin2012novel, chun2015direct}, $\alpha$-Li$_2$IrO$_3$~\cite{singh2012relevance, williams2016incommensurate} , H$_3$LiIr$_2$O$_6$~\cite{kitagawa2018spin}, and Ag$_3$LiIr$_2$O$_6$~\cite{bahrami2019thermodynamic}, as well as $\alpha$-RuCl$_3$~\cite{plumb2014spin, sandilands2015scattering, sears2015magnetic, majumder2015anisotropic, johnson2015monoclinic, sandilands2016spin, banerjee2016proximate, banerjee2017neutron, do2017majorana}. Although several material candidates have been identified, diagnosing spin-liquid states through conventional experimental probes, such as inelastic neutron scattering or thermodynamic measurements, remains extremely challenging. Attempts to detect a chiral KSL state in $\alpha$-RuCl$_3$ via the predicted quantization of the thermal Hall effect have also been inconclusive, mainly due to sample dependence and the difficulty of disentangling magnetic and phonon contributions to the thermal Hall conductivity~\cite{Kasahara18,Yamashita20,Tokoi21,Lefran22,Kasahara22,Bruin22,Czajka23,Zhang23,Kumpei24}. Hence, a natural question to ask is whether alternative experimental probes could provide more definitive evidence of this novel state of matter.

Inelastic electron tunneling spectroscopy~\cite{Lambe68,Tsui71} has been recently identified as an effective probe of magnetic excitations in two-dimensional insulators~\cite{Klein18,Ghazaryan18,Kim19}. It has also been successfully applied to resolve spin waves in interacting magnetic atoms~\cite{Balashov06,Spinelli14}, and could even offer insights into localized boundary modes~\cite{Delgado13,Costa20,Yin20}. This spectroscopic technique uses the magnetic insulator as a tunneling barrier between two metallic electrodes, and measures inelastic tunneling processes where an electron tunneling through the magnetic insulator creates spin excitations. In the context of Kitaev materials, a recent experimental work employed this technique to probe the magnetic excitations of atomically thin $\alpha$-RuCl$_3$~\cite{Yang23}. At the same time, several theoretical studies have proposed a local version of this technique, corresponding to inelastic scanning tunneling microscopy (STM), for observing key hallmarks of chiral KSLs, including the chiral Majorana edge modes~\cite{Feldmeier20} and the Majorana zero modes~\cite{Konig20,Bauer23,Takahashi23,Kao24a,Kao24b}.

In particular, Feldmeier \emph{et al.}~\cite{Feldmeier20} proposed to use such an inelastic STM setup [see Fig.~\ref{fig:dangling}(b)] in order to probe the local spin dynamics in a chiral KSL and detect gapless spin excitations along its boundary. From the inelastic STM measurement, the local dynamical spin structure factor (DSSF) can be extracted as the derivative of the tunneling conductance $dI/dV$ with respect to the bias voltage $V$. More precisely,
\begin{eqnarray}
    {dI \over dV} = \frac{2e^2 }{ \hslash} \sum_{j,l,\mu} T({\bm r}-{\bm r}_j)T({\bm r}-{\bm r}_l) \int_0^{eV} d \omega \, S^{\mu \mu}_{jl}(\omega),
\end{eqnarray}
where $S^{\mu \mu}_{jl}(\omega) = 
(2\pi)^{-1} \int_{-\infty}^{+\infty} dt \, e^{i \omega t} \langle \sigma_j^{\mu} (t) \sigma_l^{\mu} (0) \rangle$ is the real-space DSSF, and $T({\bm r}-{\bm r}_j)$ is the electron tunneling amplitude between the substrate and the STM tip at position ${\bm r}$ through site $j$ of the spin system in between [see Fig.~\ref{fig:dangling}(b)]. Since this tunneling amplitude  decays exponentially on the nanometer scale, this approach allows the detection of the local DSSF with a spatial resolution close to the atomic size.


As explained in Ref.~\cite{Feldmeier20}, inelastic STM can be employed to measure the gapless spin excitations near the boundary, which are in stark contrast to the gapped spin excitations in the bulk of the system. Since the corresponding gapless edge mode is necessarily of fermionic character, it is directly indicative of spin fractionalization and thus reveals an underlying spin liquid. At the same time, this kind of inelastic STM measurement along a clean boundary is insufficient to diagnose the chiral nature of the spin liquid because it cannot distinguish between chiral and helical gapless edge modes.


Here, we propose a definitive STM signature of chiral KSLs that is revealed by ubiquitous edge disorder but ultimately originates from ``dangling'' Majorana fermions of the underlying Kitaev model. These dangling edge fermions are directly coupled to the chiral edge fermions by the external magnetic field that produces the chiral KSL by gapping out the bulk fermion spectrum~\cite{Takahashi23, Kao24a, Kao24b, Kao21}. Hence, this significant (but generally overlooked) effect modifies the spin dynamics along the edge to linear order in the field and gives rise to a range of observable consequences. In the companion paper~\cite{comp}, we establish that a zigzag edge of the KSL supports a pronounced low-energy peak in the local DSSF whose energy scales linearly with only one component of the field. In this paper, we demonstrate that edge disorder generically produces additional resonance peaks in the local DSSF that exhibit the same linear energy scaling but can also be leveraged to diagnose the chiral nature of the KSL. In particular, the hybridization of the dangling fermions and the chiral edge mode creates a topological obstruction to Anderson localization~\cite{Altland15,Micklitz24,Park24}, which is not present in non-chiral KSLs supporting helical edge modes. Consequently, the peak width scales linearly with the field in chiral KSLs, while it is largely field independent for non-chiral KSLs. From a more general perspective, the dangling fermions act as ``witnesses'' to the topologically protected chiral edge mode, providing a unique perspective on the chiral nature of the bulk spin liquid.



\section{Model}

To study the boundary modes of the KSL under an external field, we consider the following Hamiltonian:
\begin{eqnarray}
    \hat{\mathscr H} = \hat{\mathscr H}_{K} + \hat{\mathscr H}_{G} + \hat{\mathscr H}_{Z},
\end{eqnarray}
where
\begin{equation}
    \hat{\mathscr H}_K =K_x \sum_{\langle jl\rangle \parallel x} \sigma_j^x \sigma_l^x + K_y \sum_{\langle jl\rangle \parallel y} \sigma_j^y \sigma_l^y + K_z \sum_{\langle jl\rangle \parallel z} \sigma_j^z \sigma_l^z
\end{equation}
is the exactly solvable Kitaev model~\cite{Kitaev06},
\begin{eqnarray}
    \hat{\mathscr H}_{G} &=& J \sum_{\alpha}\sum_{\langle j l \rangle} \sigma_j^{\alpha} \sigma_l^\alpha + \Gamma \sum_{\langle jl \rangle \parallel \alpha} \ \ \sum_{\beta\neq \gamma \neq \alpha} \sigma_j^\beta \sigma_l^\gamma\nonumber \\
    &+& \Gamma^\prime \sum_{\langle jl \rangle \parallel \alpha} \ \ \sum_{\beta\neq \alpha} \left( \sigma_j^\alpha \sigma_l^\beta + \sigma_j^\beta \sigma_l^\alpha \right)
\end{eqnarray}
refers to generic non-Kitaev symmetric spin-exchange interactions, and
\begin{eqnarray}
\hat{\mathscr H}_{Z} =\sum_j (h_x \sigma_j^x +h_y \sigma_j^y +h_z \sigma_j^z)
    \label{eq:Hamil}
\end{eqnarray}
represents a Zeeman coupling to an external magnetic field ${\bm h}$. Here $\langle jl \rangle$ denotes nearest neighbor bonds with three different orientations labeled by $\alpha=x,y,z$, while $h_\mu$ ($\mu=x,y,z$) are the components of the external magnetic field and $K_{\alpha}$ ($\alpha=x,y,z$) are antiferromagnetic Kitaev spin exchange interactions. This model is exactly solvable in absence of magnetic field (${\bm h} = 0$) and non-Kitaev interactions ($J=\Gamma=\Gamma^\prime = 0$), giving rise to a  quantum spin liquid ground state with two types of elementary excitations. One type corresponds to gapped or gapless matter fermions, depending the anisotropy of the Kitaev interactions, while the other one corresponds to emergent fluxes. A single flux excitation is also known as ``vison'' and has a finite energy gap $\Delta_{\rm v}$. The KSL phase is stable for a finite range of magnetic fields and non-Kitaev interactions~\cite{Zhang21}, although a magnetic field immediately gaps out any gapless matter fermions. 

\begin{figure}
    \centering
    \includegraphics[width=0.5\textwidth]{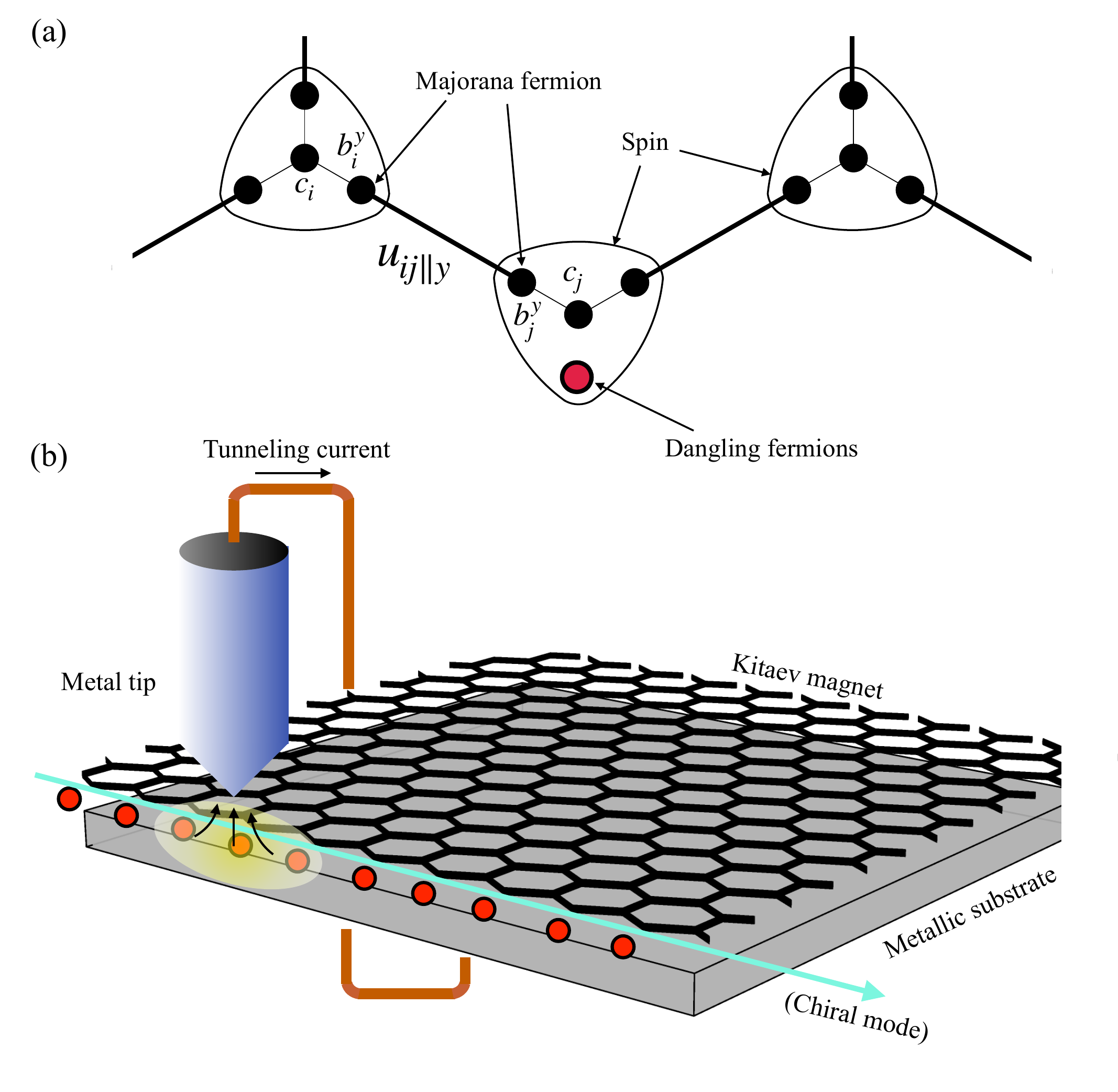}
    \caption{(a) Representation of the spin operators in terms of Majorana fermions (dots) and illustration of a dangling fermion (red dot).
    (b) Schematic STM setup for implementing spin spectroscopy on the boundary of a 2D magnet.}
    \label{fig:dangling}
\end{figure}

As illustrated in Fig.~\ref{fig:dangling},  the KSL is obtained by expressing the spin operator as a bilinear product of Majorana fermion operators~\cite{Kitaev06}, $\sigma_j^\mu = i b_{j}^\mu c_j$, where $\mu = x,y,z$, and $j$ labels the lattice site. In general, the $c$ fermions correspond to matter-field excitations, while the $b$ fermions describe gauge fields coupled to them. For an infinite lattice, the $b$ fermions pair up and form the $Z_2$ gauge fields $u_{jl\parallel \mu} = i b_j^\mu b_{l}^\mu = \pm 1$, which are conserved quantities and take uniform values $+1$ (assuming $j\in A$ and $l\in B$, with $A,B$ being the two sublattices of the honeycomb lattice) in the standard gauge choice for the zero-flux sector. For a finite lattice, however, the Z$_2$ gauge fields $u_{jl}$ are no longer well defined for the missing bonds on the boundary.  These unpaired $b$ fermions associated with missing bonds, as indicated by red dots in Fig.~\ref{fig:dangling}, are referred to as ``dangling fermions''~\cite{Takahashi23, Kao24a, Kao24b, Kao21}. The dangling fermions correspond to zero-energy boundary modes of the pure Kitaev model and must be included in the low-energy description of the spin dynamics on the boundary.


\subsection{Effective Hamiltonian under field}
\label{sec:model_eff}

In this work, we are only interested in low-energy dynamics below $\Delta_{\rm v}$, which corresponds to the zero-flux sector of the Kitaev model. In this regime, the spin dynamics is completely determined by the spin excitations generated by the ``dangling spins'' on the boundary that create ``dangling fermions''. For that purpose, we first introduce a relevant effective Hamiltonian governing the low-energy dynamics.

\begin{figure}
    \centering
    \includegraphics[width=0.5\textwidth]{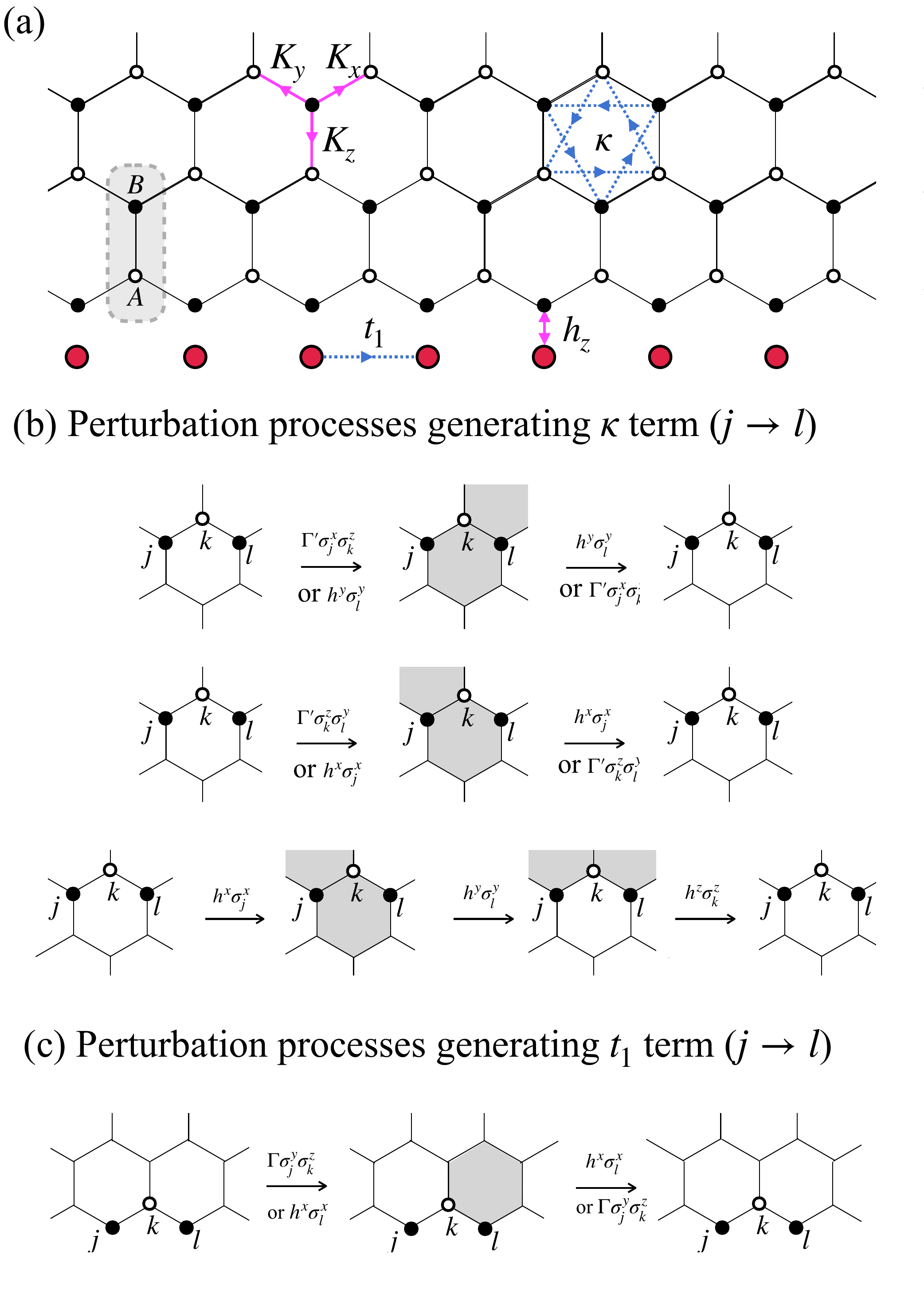}
    \caption{(a) Honeycomb lattice with a perfect zigzag boundary that contains a sequence of dangling fermions (red dots). The gray rectangle marks the unit cell of the honeycomb lattice with $A$ and $B$ sublattices. Pink and blue arrows correspond to quadratic terms in the effective Majorana-fermion Hamiltonian of Eq.~\eqref{eq:Hmajorana}. 
    (b) Second-order processes (first two lines) and third-order processes (last line) that result in the $\kappa$ term in Eq.~\eqref{eq:kappa}. Note that permutations of $(\sigma_j^x,\sigma_k^z,\sigma_l^y)$ leading to the same matrix elements are not shown here. (c) Second-order processes giving rise to the $t_1$ term in Eq.~\eqref{eq:t}.}
    \label{fig:boundary}
\end{figure}

Projecting $\hat{\mathscr H}$ into the zero-flux sector  yields two distinct contributions. The first one originates from the Kitaev Hamiltonian $\hat{\mathscr H}_K$ and has a quadratic form in terms of the $c$ fermions:  
\begin{equation}
\hat{\tilde{\mathscr H}}_K = i \sum_{\langle jl\rangle} \hat{c}_{j\in A} \hat{c}_{l\in B},
\end{equation}
where the sum runs over all non-equivalent nearest neighbor bonds, $A$ and $B$ refer to the two sublattices of the honeycomb lattice, and we are adopting the standard gauge choice for the zero-flux sector~\cite{Kitaev06}. The second contribution arises from the Zeeman coupling to an external field, $\hat{\mathscr H}_Z$.  Projected into the zero-flux sector, this term generates a boundary contribution, 
\begin{equation}
\label{eq:Hz1}
\hat{\tilde{\mathscr H}}_Z = i \sum_{j\in {\rm boundary}} h^{\mu_j} \hat{b}_j^{\mu_j}\hat{c}_j,
\end{equation}
which is \emph{linear} in the magnetic field and hybridizes each dangling fermion $\hat{b}_j^{\mu_j}$ (where $\mu_j$ is $x$, $y$, or $z$, depending on $j$) with the matter fermion $\hat{c}_j$ via the $\mu_j$-th component of the Zeeman field.  The two terms described in this paragraph are illustrated by the pink solid arrows in Fig.~\ref{fig:boundary}(a).

Higher-order processes generate further symmetry-allowed terms, as indicated by blue dashed arrows in Fig.~\ref{fig:boundary}(a). Given the dominance of the Kitaev interactions, we restrict our attention to terms arising from leading-order perturbations. Additionally, we assume that these interactions do not alter the quasiparticle picture, retaining only quadratic terms in the effective Hamiltonian. In the bulk, higher-order processes generate the effective three-spin interaction that immediately gaps out any gapless bulk fermions~\cite{Kitaev06}:
\begin{equation}\label{eq:kappa}
\hat{\tilde{\mathscr H}}_{1}^{\prime} = - \kappa \sum_{\langle jkl\rangle_{\alpha \beta \gamma}} \sigma_j^{\alpha} \sigma_k^{\beta} \sigma_l^{\gamma} = i \kappa \sum_{\langle jkl\rangle_{\alpha \beta \gamma}} u_{kj \parallel \alpha}u_{kl \parallel \gamma}c_j c_l,
\end{equation}
where $(\alpha\beta\gamma)$ is a permutation of $(xyz)$ in each term, and $\langle jkl\rangle_{\alpha \beta \gamma}$ denotes a three-site path formed by two consecutive nearest-neighbor bonds $\langle jl\rangle \parallel \alpha$ and $\langle lk\rangle \parallel \beta$. This effective interaction preserves the integrability of the model. The scaling of $\kappa$ with respect to ${\bm h}$ is given by $c_1 h + c_2 h^3 + \dots$ \cite{Song16}, where $c_1 \sim 4\Gamma^\prime/\Delta_{\rm v}$ originates from a second-order process involving the $\Gamma^\prime$ term the and Zeeman coupling term, while $c_2$ arises from a third-order contribution in ${\bm h}$~\cite{Kitaev06}. The factor of $4$ in $c_1$ accounts for two different processes, illustrated by the first two lines of Fig.~\ref{fig:boundary}(b), which contribute to the same matrix element. On the boundary, higher-order processes endow the dangling fermions with a finite hopping term:
\begin{equation}\label{eq:t}
\hat{\tilde{\mathscr H}}_{2}^{\prime} = \sum_{jl \in {\rm boundary}} t_{jl} b_j^{\mu_j} b_l^{\mu_l}.
\end{equation}
This hopping arises, for example, from second-order perturbation processes involving the $\Gamma$ term and the Zeeman coupling term, as shown in Fig.~\ref{fig:boundary}(c), which give a hopping amplitude $t_{jl} \sim 4h\Gamma/\Delta_{\rm v}$. In the following, we consider only the nearest-neighbor hopping term, denoted as $t_1$. Hopping terms beyond nearest neighbors merely affect the dispersion of the boundary mode without introducing any qualitatively new effects.

To summarize, the low-energy physics is described by the effective Hamiltonian 
\begin{equation}
\hat{\tilde{\mathscr H}} = \hat{\tilde{\mathscr H}}_K + \hat{\tilde{\mathscr H}}_Z + \hat{\tilde{\mathscr H}}_1^{\prime} + \hat{\tilde{\mathscr H}}_2^{\prime} + ...,
\end{equation}
which takes a quadratic form,
\begin{equation} \label{eq:Hmajorana}
    \hat{\tilde{\mathscr H}} = i \sum_{mn} A_{mn}\hat{F}_m  \hat{F}_n,
\end{equation}
in the fermionic representation. Here $A_{mn} = -A_{nm}\in {\cal R}$, and $\hat{F}_m$ refers to the components of the Majorana-fermion vector,
\begin{eqnarray}\label{eq:mfopr}
    \hat{\bm F} = (\hat{c}_1, ..., \hat{c}_N, \hat{b}_1^{\mu_1}, ..., \hat{b}_M^{\mu_M})^T,
\end{eqnarray}
with $N$ being the number of lattice sites, $M$ the number of dangling fermions on the boundary, and $\mu_j$ the spin component corresponding to a given dangling fermion (see, e.g., Fig.~\ref{fig:boundary}).

The effective Hamiltonian $\hat{\tilde{\mathscr H}}$ can be diagonalized through a unitary transformation, 
\begin{eqnarray}\label{eq:transf}
\hat{F}_m =  \sqrt{2} \sum_{s=1}^{\cal N} \Phi_{m s} \hat{f}_s + {\rm H. c.},
\end{eqnarray}
where ${\cal N}= N +M$, and ${\bf \Phi}$ is an ${\cal N} \times {\cal N}$ unitary matrix.  The $s$-th column of this matrix, ${\bm \Phi}_s$, satisfies the eigenvalue equation 
\begin{eqnarray}
    i {\bm A} {\bm \Phi}_{s} = \epsilon_s {\bm \Phi}_s, \ \ i {\bm A} {\bm \Phi}_s^* = -\epsilon_s {\bm \Phi}_s^*,
\end{eqnarray}
where ${\bm A}$ is  the matrix of  the quadratic Hamiltonian $\hat{\tilde{\mathscr H}}$ in Eq.~\eqref{eq:Hmajorana}. The eigenvalues appear in pairs $\pm \epsilon_s$ due to particle-hole symmetry. After applying the transformation in Eq.~\eqref{eq:transf}, the effective Hamiltonian takes a diagonal form
\begin{equation}
    \hat{\tilde{\mathscr{H}}} = \sum_{\epsilon_s \geq 0} \epsilon_s \hat{f}_s^\dagger \hat{f}_s.
\end{equation}
To distinguish the $b$ and $c$ fermions, we introduce the  notations $\hat{\bm c}\equiv (\hat{c}_1, ..., \hat{c}_N)^T$ and $\hat{\bm b}\equiv (\hat{b}_1^{\mu_1},..., \hat{b}_M^{\mu_M})^T$. Moreover, we define ${\bm \Phi}_s^T \equiv ({\bm \varphi}_s^T, {\bm \phi}_s^T)$, where ${\bm \varphi}_s$ is a $N\times 1$  vector and ${\bm \phi}_s$ a $M\times 1$ vector. The transformation in Eq.~(\ref{eq:transf}) leads to two equations
\begin{eqnarray}
\label{eq:transfbc}
\hat{c}_j &=& \sqrt{2} \sum_s \varphi_{js} \hat{f}_s + {\rm H. c.}, 
\nonumber \\
\hat{b}_j^{\mu_j} &=& \sqrt{2} \sum_s \phi_{js} \hat{f}_s + {\rm H. c.},
\end{eqnarray}
where $\varphi_{js}$ and $\phi_{js}$ are the $j$-th components of ${\bm \varphi}_s$ and ${\bm \phi}_s$, respectively. The inverse transformation is given by
\begin{eqnarray}
\hat{f}_s^\dagger = {1\over\sqrt{2}} \left( \sum_{j=1}^N \varphi_{js} \hat{c}_j+ \sum_{j=1}^M \phi_{js}  \hat{b}_j^{\mu_j}\right). 
\end{eqnarray}


\subsection{Local spin structure factor}

Our primary focus  is  the local dynamic spin structure factor (DSSF) on  boundary lattice sites $j$,
\begin{eqnarray}
    {\cal S}_{j}^{\mu_j \mu_j} (\omega) = {1\over 2\pi}\int_{-\infty}^{\infty} dt e^{i\omega t} \langle \sigma_j^{\mu_j} (t) \sigma_j^{\mu_j} (0) \rangle,
\end{eqnarray}
which can be  measured by STM. We specifically consider the spin components $\mu_j$ corresponding to the dangling fermions at the boundary sites $j$, as they excite low-energy boundary fermion modes without generating $Z_2$ gauge fluxes with a finite energy gap $\sim \Delta_{\rm v}$. 

In the Majorana fermion representation, the local DSSF 
reduces to a convolution of two Majorana fermion  spectral functions ${\cal A}_{F_m F_m} (\omega)$:
\begin{eqnarray}\label{eq:local}
    S_j^{\mu_j\mu_j}(\omega) &=& 
    \int_0^\omega d{\nu} \left[{\cal A}_{b_j, b_j}(\nu) {\cal A}_{c_j, c_j}(\omega-\nu) \right. \nonumber \\ 
   && \left. -{\cal A}_{b_j, c_j}(\nu) {\cal A}_{c_j, b_j}(\omega-\nu)\right],
\end{eqnarray}
where we have abbreviated the flavor index of the dangling fermions for notation convenience, and  
\begin{eqnarray}\label{eq:Aw_def}
    {\cal A}_{F_m F_n} (\omega) = {1\over 2\pi}\int_{-\infty}^{\infty} dt e^{i\omega t} \langle \hat{F}_m (t) \hat{F}_n (0) \rangle.
\end{eqnarray}
Since $\hat{F}^2_{m} =1$, the spectral function  satisfies the sum rule 
\begin{equation}
\int_{-\infty}^{\infty} d\omega {\cal A}_{F_m F_m}(\omega) = 1,
\end{equation}
while the off-diagonal elements satisfy 
\begin{equation}
\int_{-\infty}^{\infty} d\omega {\cal A}_{F_m F_n}(\omega) = - \int d\omega {\cal A}_{F_n F_m}(\omega)
\end{equation}
due to the anti-commutation relation of the Majorana fermions: $\{ \hat{F}_m, \hat{F}_n \} = 2\delta_{mn}$. Furthermore, since  $\sigma_j^{\mu} \sigma_j^{\mu} = \sigma_j^0$, the DSSF satisfies the sum rule  
\begin{eqnarray}
    \int_{-\infty}^{\infty} d\omega S_j^{\mu_j\mu_j}(\omega) = 1.
\end{eqnarray}
The Majorana fermion spectral functions can be computed by expressing the $\hat{F}_m$ operators in terms of the quasi-particle operators $\hat{f}_s$ that diagonalize  $\tilde{\mathscr H}$:
\begin{eqnarray}\label{eq:Aw}
    {\cal A}_{F_m F_n} (\omega) &=& \sum_{\epsilon_s \geq 0} \left[  {\cal M}^*_{F_m}(s){\cal M}_{F_n}(s) \delta(\omega-\epsilon_s) \right. \nonumber \\
    &+& \left.   {\cal M}^*_{F_n}(s){\cal M}_{F_m}(s)  \delta(\omega+\epsilon_s) \right],
\end{eqnarray}
where
\begin{eqnarray}\label{eq:matrix_element}
    {\cal M}_{F_m} (s) = \langle s \rvert \hat{F}_m \rvert0\rangle = \sqrt{2} \Phi_{m s}^*,
\end{eqnarray}
and $\rvert s \rangle = \hat{f}_s^\dagger \rvert0\rangle$ represents an excited 
single-quasiparticle state.

\section{ STM response for a uniform boundary }
\label{sec:clean_boundary}

A distinctive feature of the chiral KSL, setting it apart from the non-chiral KSL as well as conventional ordered magnets, is the existence of a gapless chiral boundary mode. Detecting this chiral boundary mode is essential for experimentally identifying this exotic state of matter. In the companion paper~\cite{comp}, we studied the local DSSF of the KSL, a quantity accessible by STM, and its field dependence. Specifically, we demonstrated that the hybridization between the dangling fermions and the gapless boundary mode leads to a pronounced peak in the local DSSF whose energy scales linearly with only one component of the magnetic field.

The goal of this work is to explore how STM can be used to distinguish the chiral KSL from the non-chiral one, which lacks chiral boundary modes, and more broadly, from other magnetic states with low-energy boundary modes. To achieve this goal, we proceed in two steps. In this section, we consider a clean and uniform boundary by adopting periodic boundary conditions (PBC) along the boundary direction (cylindrical lattice) and compare the local DSSFs for the chiral and non-chiral KSL phases. Given that the distinction between the chiral and non-chiral KSLs is inevitably complicated by disorder, we address this important effect in the next section, highlighting how boundary defects can be used to detect universal signatures of the chiral boundary mode.

\begin{figure*}
    \centering
    \includegraphics[width=0.8\textwidth]{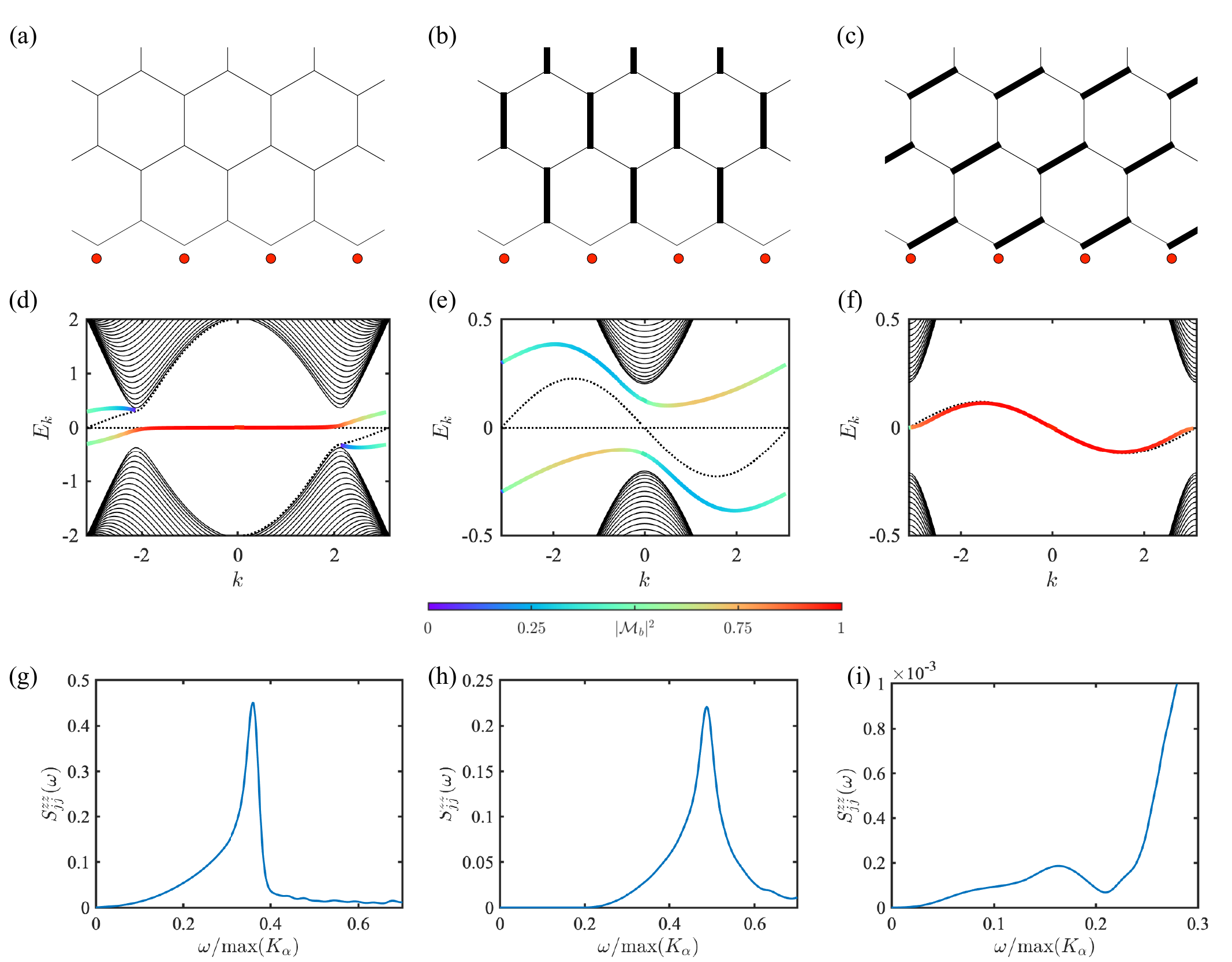}
    \caption{(a)-(c) Zigzag boundaries for (a) the isotropic Kitaev model with $K_x=K_y=K_z=1$, $h^z=0.15$, $\kappa=0.2h^z$, $t_1=0$, (b) the anisotropic Kitaev model with $K_x=K_y=0.45,K_z=1$, $h^z=0.15$, $\kappa=0.2h^z$, $t_1 = 0$, and (c) the anisotropic Kitaev model with $K_y=K_z=0.45,K_x=1$, $h^z=0.15$, $\kappa=0.2h^z$, $t_1 = 0.4h^z$. (d)-(f) Corresponding mode dispersions along the boundary. Colored solid lines (black dotted lines) mark boundary modes with (without) the hybridization term $\hat{\tilde{\mathscr{H}}}_{Z}$ in Eq.~(\ref{eq:Hz1}), with the color scale indicating the degree of dangling-fermion character. (g)-(e) Corresponding local DSSFs at boundary sites supporting dangling fermions.}
    \label{fig:local_Sw}
\end{figure*}

\subsection{Chiral spin liquid phase}

We begin by generalizing the analysis in Ref.~\cite{comp} to arbitrary Kitaev spin exchange. 
Since we are adopting PBC along the boundary direction, the eigenmodes have a well defined quasi-momentum $k=2\pi n /L$,  where $n$ is an integer number that belongs to the interval  $-L/2 < n \leq L/2$ ($L$ is assumed to be even). For a generic applied magnetic field with $\kappa \neq 0$, the energy spectrum consists of gapped bulk modes, where a gapless Dirac cone at $k=k_0$ acquires a finite mass $\propto \kappa$. 

In the absence of the hybridization term  $\hat{\tilde{\mathscr{H}}}_{Z}$, the boundary modes consist of two branches: one branch features a flat zero-energy spectrum representing the dangling fermions, while the other branch corresponds to a chiral mode formed by the matter fermions. This chiral mode only exists for $\rvert k \rvert > k_0$ and connects the negative-energy Dirac cone at $k=-k_0$ with the positive-energy Dirac cone at $k=k_0$. 

The Zeeman term strongly  hybridizes the dangling fermion modes at $\rvert k \rvert > k_0$  with the chiral mode, while those at $\rvert k \rvert < k_0$ retain their dominant dangling-fermion character because they can only hybridize with bulk matter-fermion modes. Fig.~\ref{fig:local_Sw}(d)  shows the boundary modes on one edge before (black dashed line) and after (solid colored line) hybridization for the zigzag boundary depicted in Fig.~\ref{fig:local_Sw}(a). 


Fig.~\ref{fig:local_Sw}(a,d,g) show the results obtained from  numerical diagonalization of the effective Hamiltonian in Eq.~\eqref{eq:Hmajorana}. While these calculations are performed for a finite value  of the magnetic field, the general structure can be understood by analyzing the small-field limit, $\rvert h_\mu \rvert \ll \Delta_{\rm v}$, where the spectrum of the boundary modes can be found analytically. 
For simplicity, we also consider the limit of $\kappa =0$. An analytical calculation reveals that the crossover momentum $k_0$ is then given by
\begin{eqnarray}\label{eq:k0}
    k_0 = \cos^{-1} \left( \frac{K_z^2 - K_x^2 - K_y^2}{2 K_x K_y} \right),
\end{eqnarray}
which recovers $k_0 = 2\pi/3$ in the isotropic limit. The condition $0<k_0<\pi$ holds within the entire non-Abelian phase.

For $\rvert k\rvert<k_0$, the boundary-mode energy is zero to linear order in $h_\mu$ and has dominant dangling-fermion character, as captured by the wave function
\begin{eqnarray}\label{eq:wf_flat}
    \varphi_{j,k} \simeq 0, \qquad \phi_{j,k} \simeq \frac{1}{\sqrt{L}} e^{i k j}.
\end{eqnarray}
Note that Majorana fermions are real operators, implying that $b_{-k}^\dagger = b_k$ describe the same degree of freedom.

For $\rvert k\rvert>k_0$, the dangling and matter fermions form a hybridized mode with energy linear in $h_\mu$. For the zigzag boundary orientation shown in Fig.~\ref{fig:local_Sw}(a), this hybridization is linear in the 
$z$-component $h_z$ of the external field~\cite{comp}:
\begin{eqnarray}\label{eq:ek}
    E_k = 2h_z \sqrt{1-\lambda_k^2},
\end{eqnarray}
where
\begin{eqnarray}\label{eq:lambda}
    \lambda_k^2 = \frac{K_x^2 + K_y^2 + 2 K_x K_y \cos k}{K_z^2}
\end{eqnarray}
is a dimensionless parameter defined by $\lambda_k = \exp(-9/4 \xi_k)$, with $\xi_k$ being the the localization length of the boundary mode for wave vector $k$. Clearly, the boundary mode only exists for $\lambda_k^2 < 1$, and  the crossover momentum $k=k_0$ given in Eq.~\eqref{eq:k0} precisely corresponds to $\lambda_k^2 = 1$.


The minimum of $E_k$ is at $k=k_0$ where $E_k=0$. Around this point, the dispersion reads
\begin{eqnarray}\label{eq:ek_squareroot}
    E_k \simeq 2h_z \left[ \left(\frac{K_x-K_y}{K_z}\right)^2+1 \right]^{1/4} 
    \left[ \left(\frac{K_x+K_y}{K_z}\right)^2-1 \right]^{1/4} 
     \sqrt{\rvert k\rvert-k_0}.
     \nonumber \\
\end{eqnarray}
The singular square-root behavior of the dispersion leads to the following density of states (DOS) for this hybridized mode, 
\begin{eqnarray}
    \rho(\omega) &=& \frac{1}{\pi}\int_{k_0}^\pi dk \delta(\omega-E_k) =
    \frac{1}{\pi} \frac{1}{\rvert {d E_k / dk}\rvert } \rvert_{E_k = \omega},
\end{eqnarray}
which is linear in $\omega$ for small energies: $\rho(\omega) \propto \omega$. The maximum of $E_k$ is at $k=\pi$ where $E_{\pi} = 2h_z \sqrt{1-(K_x-K_y)^2/K_z^2}$. Correspondingly, the DOS has an inverse square-root divergence at the top of the band: $\rho(\omega) \propto 1/\sqrt{E_\pi - \omega}$.

Next, we investigate the local DSSF, which requires us to compute the matrix elements defined in Eq.~\eqref{eq:matrix_element}. For momenta $0< \rvert k\rvert < k_0$, we have
\begin{eqnarray}\label{eq:m1}
    {\cal M}_{b_j} (k) \simeq \sqrt{\frac{2}{L}} e^{-i k j}, \ \ {\cal M}_{c_j} (k) \simeq 0,
\end{eqnarray}
while for $k_0 < \rvert k\rvert < \pi$, we obtain
\begin{eqnarray}
    {\cal M}_{b_j} (k) &\simeq& \frac{i}{\sqrt{L}} e^{-i k j + i \phi_k/2}, \label{eq:m2} \\
    {\cal M}_{c_j} (k) &\simeq& \frac{1}{\sqrt{L}} \frac{E_k}{2 h_z} e^{-i k j + i \phi_k/2}, \label{eq:m3}
\end{eqnarray}
where $\phi_k$ is a $k$-dependent phase factor. Notably, these matrix elements are explicit functions of the boundary-mode energies $E_k$. From these results, we obtain the Majorana-fermion spectral functions in Eq.~\eqref{eq:Aw}:
\begin{eqnarray}
    {\cal A}_{bb}(\omega) &=& \frac{2}{3} \delta(\omega) + \rho(\omega), \label{eq:Abb} \\
    {\cal A}_{cc}(\omega) &=& \left(\frac{\omega}{2h_z}\right)^2 \rho(\omega),\label{eq:Acc} \\
    {\cal A}_{bc}(\omega) &=& -{\cal A}_{cb}(\omega) =  i\left(\frac{\omega}{2h_z}\right) \rho(\omega).\label{eq:Abc}
\end{eqnarray}
One can verify that $\int_{-\infty}^{\infty} d\omega {\cal A}_{bb}(\omega) = 1$, obeying the sum rule. In contrast, the integrated spectral weight of  ${\cal A}_{cc}(\omega)$ is less than 1, as some  weight is transferred  to the bulk modes. The $\delta$-peak in ${\cal A}_{bb}$ arises from the quasi-flat modes at $\rvert k \rvert <k_0$, which, in reality, have a finite broadening $\propto \kappa$ that is neglected in the small-field analysis above. The factors $\omega/(2h_z)$ in ${\cal A}_{cc}$, ${\cal A}_{bc}$, and ${\cal A}_{cb}$ originate from the factor $E_k/(2h_z)$ in the matrix element ${\cal M}_{c_j}$ given in Eq.~\eqref{eq:m3}. 

The local DSSF is finally obtained by performing the convolution described in  Eq.~\eqref{eq:local}:
\begin{eqnarray}\label{eq:Sw}
    S_j^{zz}(\omega) &=& 
    \frac{2}{3}
    \left( \frac{\omega}{2h_z} \right)^2 \rho(\omega)  \nonumber \\
    &+& 
\frac{1}{2} 
    \int_0^\omega d\nu \rho(\nu)\rho(\omega-\nu) \left(\frac{\omega-2\nu}{2h_z}\right)^2,
\end{eqnarray}
where the first term arises from the convolution of the $\delta$-peak in ${\cal A}_{bb}$ with ${\cal A}_{cc}$, while the second term is connected to two-particle excitations from the hybridized mode at $\rvert k \rvert > k_0$. 

Since $\rho(\omega) \propto \omega$ for small energies, the first term in Eq.~\eqref{eq:Sw} dominates over the second one, leading to 
\begin{equation}
S_j^{zz}(\omega) \propto \omega^3.
\end{equation} 
Moreover, the first term gives an inverse-square-root singularity at the top of the band, corresponding to
\begin{equation}
S_j^{zz}(\omega) \propto 1/\sqrt{E_\pi - \omega}.
\end{equation}
In contrast, the second term results in a regular continuum without any singularities in the entire energy range. 

In Fig.~\ref{fig:local_Sw}(g), we show the local DSSF for a finite magnetic field. While the low-energy $\omega^3$ behavior is rigorously satisfied, the singularity near the band top is smeared out over the interval $\rvert \omega - E_\pi \rvert \lesssim \kappa$ because the quasi-flat mode, corresponding to the $\delta$-peak in Eq.~\eqref{eq:Abb}, has a finite broadening $\propto \kappa$. Nevertheless, the inverse-square-root behavior still approximately applies in a finite energy range below this interval.

\subsection{Non-chiral spin liquid phase}

Next, we discuss the non-chiral phase of the Kitaev model, where the Kitaev spin exchange is stronger along one of the three bond directions. As shown in Figs.~\ref{fig:local_Sw}(b) and \ref{fig:local_Sw}(c), there are two possible scenarios depending on the orientation of the strongest bonds relative to the boundary.
These two scenarios exhibit distinct boundary excitation spectra, which can be understood by examining the spectrum in the absence of the hybridization term $\hat{\tilde{\mathscr{H}}}_Z$. In the first scenario, where the strongest bonds are perpendicular to the boundary, the boundary spectrum consists of one branch of dangling fermions and another branch of matter fermions [black dotted lines in Fig.~\ref{fig:local_Sw}(e)]. Note that the existence of a matter-fermion boundary mode is linked to boundary sites that are not connected to any dominant bonds as creating a matter-fermion excitation on a dominant bond incurs an energy cost comparable to the largest Kitaev spin exchange.
In contrast, for the second scenario, where the strongest bonds are at a $30^\circ$ angle with respect to the boundary, all lattice sites belong to a dominant bond, and the boundary spectrum thus contains only a single branch of dangling fermions [black dotted line in Fig.~\ref{fig:local_Sw}(f)]. Upon introducing the hybridization term $\hat{\tilde{\mathscr{H}}}_Z$, the two boundary modes in the former case become gapped, whereas the boundary mode in the latter case remains gapless, as indicated by the colored lines in Figs.~\ref{fig:local_Sw}(e) and \ref{fig:local_Sw}(f).

\begin{figure}
    \centering
    \includegraphics[width=0.5\textwidth]{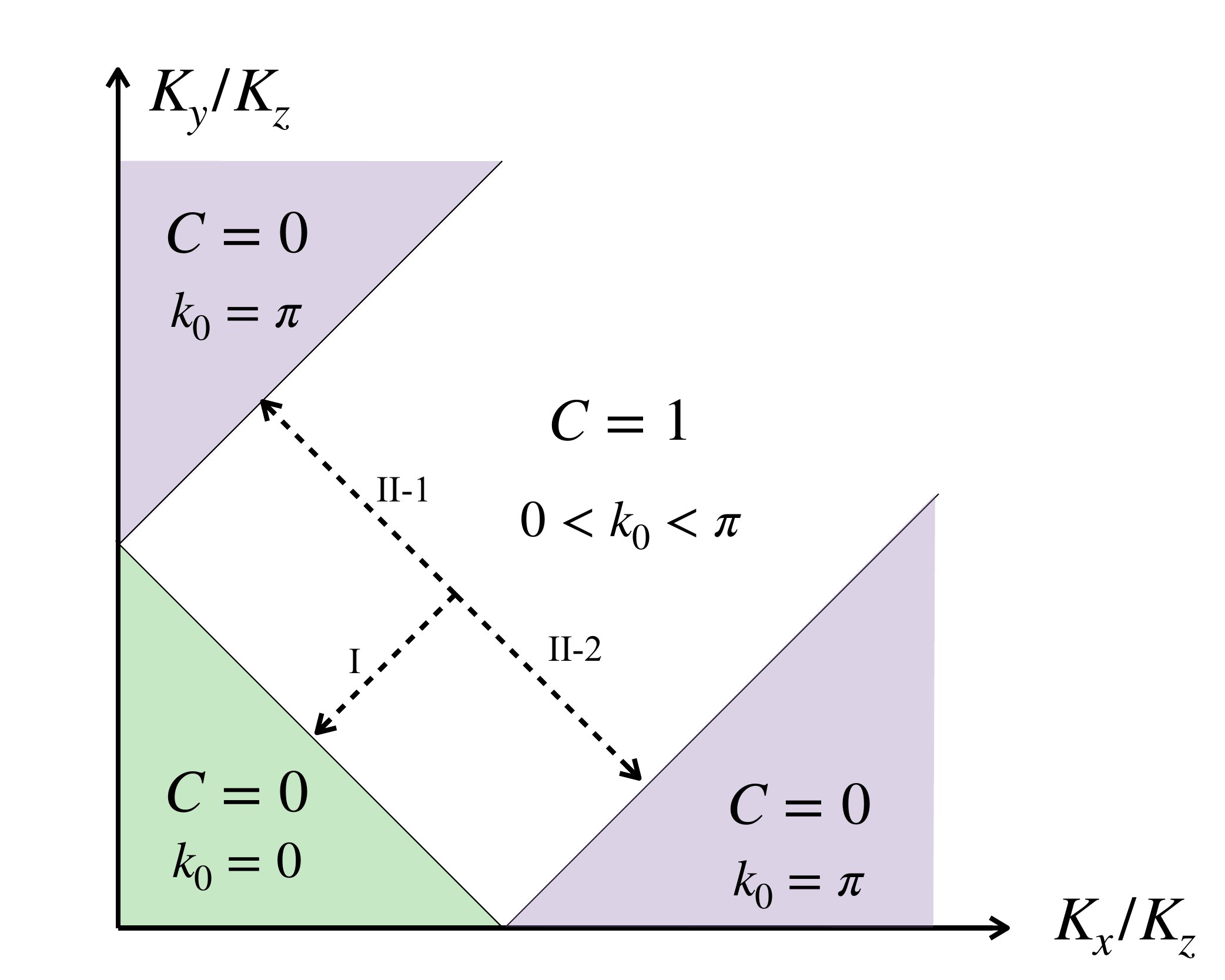}
    \caption{Non-Abelian ($C=1$) and Abelian ($C=0$) phases of the Kitaev model, along with the behavior of the crossover momentum $k_0$ in Eq.~\eqref{eq:k0}. Dashed lines indicate different paths for approaching the Abelian phase from the non-Abelian phase.}
    \label{fig:pd}
\end{figure}

To make contact with the boundary spectrum of the chiral KSL, we approach the non-chiral phase from the chiral phase by following one of the paths shown in Fig.~\ref{fig:pd}. Along path I, $k_0$ decreases monotonically and vanishes at the phase boundary into the non-chiral phase. Therefore, the dispersive hybridized mode, which exists at $\rvert k \rvert > k_0$ in the chiral KSL phase, spans the entire BZ at the phase boundary. In contrast, if we approach the non-chiral phase from path II-1 or II-2, $k_0$ increases to $\pi$ at the phase boundary, and the quasi-flat mode at $\rvert k \rvert < k_0$ extends over the whole BZ accordingly. 

In the following, we analyze the two scenarios of the non-chiral KSL separately.
We start from the scenario in Fig.~\ref{fig:local_Sw}(b) and the corresponding numerical results in Fig.~\ref{fig:local_Sw}(e,h). In this case, the analytical expressions for the small-field limit of the chiral phase remain applicable. However, they reveal different behaviors for both the  Majorana-fermion spectral function and the local DSSF. First, the energy spectrum in Eq.~\eqref{eq:ek} becomes gapped due to ${\rm max}(\lambda_k^2) = \lambda_0^2 = (K_x+K_y)^2/K_z^2 < 1$. In particular, the spectrum is bounded, $E_0 \leq E_k \leq E_\pi$, by the two finite energies
\begin{eqnarray}
    E_0 = 2h_z \sqrt{1 - (K_x+K_y)^2/K_z^2}, \\
    E_\pi = 2h_z \sqrt{1 - (K_x-K_y)^2/K_z^2}.
\end{eqnarray}
The gapped nature of the spectrum distinguishes the non-chiral phase from the chiral phase where a gapless chiral mode is guaranteed by topology. Due to the quadratic dispersions near the band bottom $E=E_0$ and the band top $E=E_\pi$, the density of states (DOS) of the boundary modes exhibits inverse-square-root singularities:
\begin{eqnarray}
\rho(\omega) \propto \frac{1}{\sqrt{\omega - E_0}},
\quad
\rho(\omega) \propto \frac{1}{\sqrt{E_\pi - \omega}}.
\end{eqnarray}
The Majorana-fermion spectral function ${\cal A}_{F_m F_n} (\omega)$ is  then confined to the energy interval $E_0 \leq \omega \leq E_\pi$. Moreover, since the matrix elements  {${\cal M}_{F_m}(k)$} in Eq.~\eqref{eq:matrix_element} remain finite in the limits of $\omega \to E_{0,\pi}$, the spectral function exhibits the same singular behavior as $\rho(\omega)$. 

The convolution of the spectral functions in Eq.~\eqref{eq:local} leads to a two-particle continuum distributed within the energy range $2E_0 < \omega < 2E_\pi$. 
While the convolution produces finite values for $S_j^{zz}({2}E_0)$ and $S_j^{zz}({ 2}E_\pi)$, a logarithmic singularity appears at $\omega = E_0 + E_\pi$:
\begin{eqnarray}
    S_j^{zz}(\omega) \propto \log \biggr \rvert \frac{1}{\omega - (E_0 + E_\pi)}  \biggr \rvert .
\end{eqnarray}
These features obtained to the leading order in $h_\mu$ explain the local DSSF presented in Fig.~\ref{fig:local_Sw}({h}).

Next, we consider the alternative scenario in Fig.~\ref{fig:local_Sw}(c), which corresponds to the numerical results in Fig.~\ref{fig:local_Sw}(f,i). 
As we discussed above, the dangling fermions can only hybridize with each other as well as the gapped bulk modes, resulting in an effective tight-binding model along the boundary. This effective tight-binding spectrum is gapless, resembling the chiral boundary mode. We also note that the effective hopping amplitude is only finite in the absence of time-reversal symmetry and is not captured by the previous small-field analysis, which only gives rise to a zero-energy mode with the wave function dominated by the dangling fermions [see Eq.~\eqref{eq:wf_flat}].



We calculate the local DSSF numerically for a finite-size lattice. As shown in Fig.~\ref{fig:local_Sw}(i), the DSSF exhibits  $\omega^3$ behavior at small energies, resembling the chiral phase. Additionally, a broad peak appears at the energy corresponding to the top of the boundary band (denoted as $E_{\rm max}$). The spectral intensity is then gradually reduced upon further increasing the energy and vanishes above $2E_{\rm max}$.


To summarize, the analysis of the chiral and non-chiral KSLs based on the effective low-energy Hamiltonian in Eq.~\eqref{eq:Hmajorana} reveals distinct spectral features in the local DSSF.  The spectrum remains gapless in the chiral KSL due to the presence of a topologically protected chiral boundary mode.
In contrast, for the non-chiral KSL, the spectral response can be either gapped or gapless, contingent on the boundary orientation relative to the dominant bond interaction.

\begin{figure*}
    \centering
    \includegraphics[width=0.9\textwidth]{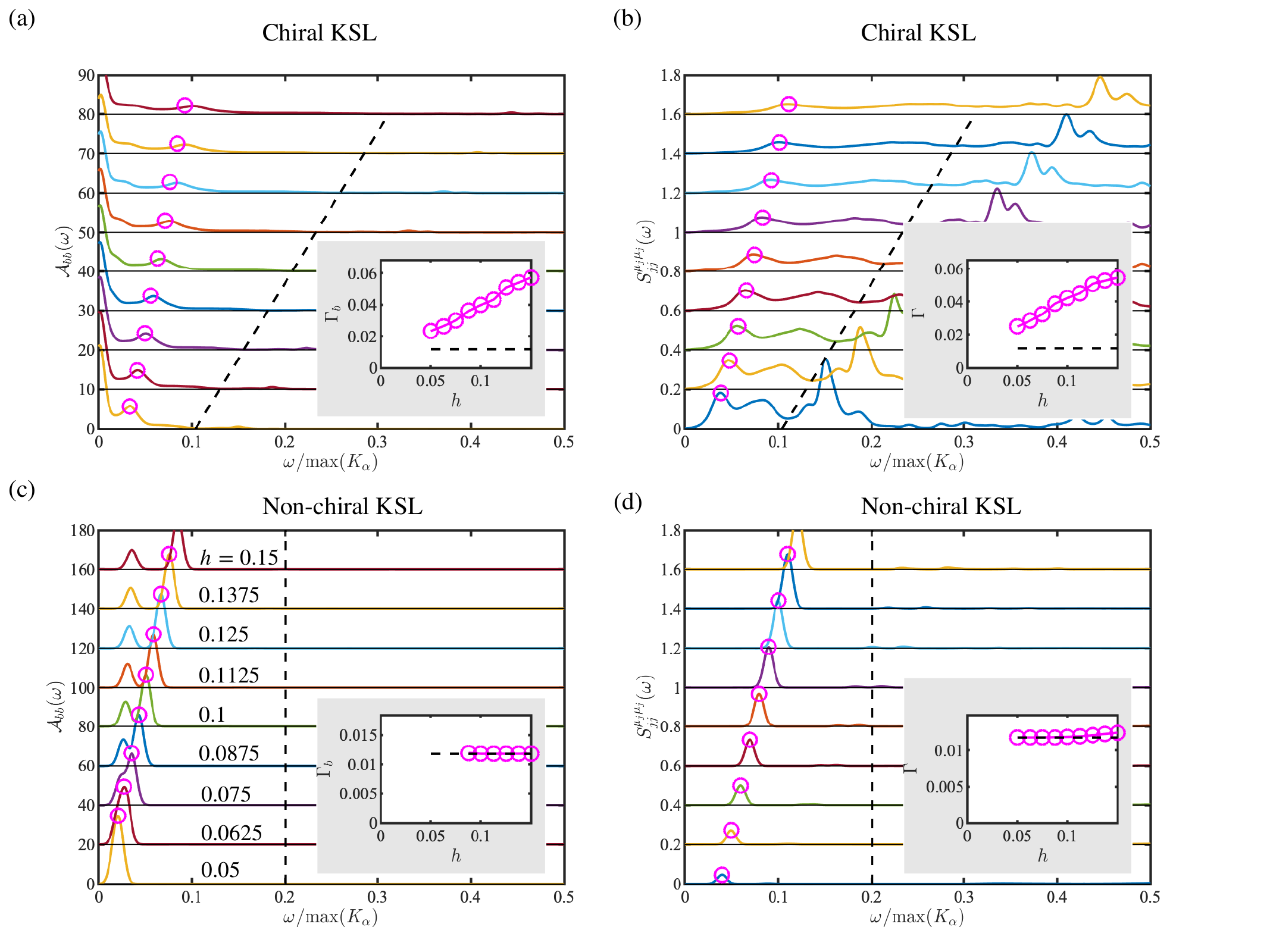}
    \caption{Local spectral function of dangling fermions and local DSSF in the presence of boundary defects. The external field is applied along the $[111]$ direction, perpendicular to the honeycomb plane: ${\bm h}=(h,h,h)$. The model parameters for the chiral and nonchiral cases are the same as in Figs.~\ref{fig:local_Sw}(a) and \ref{fig:local_Sw}(c), respectively, while the values of $h$ are indicated in panel (c). The calculation is performed on a finite lattice of $80\times 80$ unit cells, with the boundary sites cut randomly with probability $30\%$.
    The black dashed line indicates the bulk gap. Inset: half width at half maximum (HWHM) for the peak marked by pink circles in the main panel, as obtained from a Gaussian fit. The black dashed line is the energy resolution. The HWHM coincides with the energy resolution in the non-chiral case (c,d) and increases linearly with $h$ in the chiral case (a,b).} 
    \label{fig:STM_chiral_disorder}
\end{figure*}

While the characteristic lineshapes observed in both cases can serve as useful indicators, similar low-energy spectral distributions can arise in both chiral and non-chiral KSLs.  Moreover, when comparing the local DSSF for the three scenarios in Fig.~\ref{fig:local_Sw}(g,h,i), the distinction between gapped and gapless spectra can become obscured under realistic conditions due to the inevitable presence of disorder. This lack of a clear distinction motivates us to investigate the impact of boundary defects, as their presence may help uncover spectral features unique to the chiral boundary mode, distinguishing it from other types of boundary modes in, e.g., the non-chiral KSL.


\section{Effect of disorder along the boundary}
\label{sec:disorder}

Realistic boundaries lack translational invariance due to impurities, vacancies, or corners. As a result of Anderson localization~\cite{Anderson58}, non-chiral boundary modes become localized for any disorder strength. In contrast, the chiral boundary mode remains delocalized because it is protected by the bulk topology. In this section, we explore the measurable effects of this key distinction using STM.

To simulate disorder, we randomly remove lattice sites along the boundary with a finite probability $0 < p < 1$. Although we use $p = 0.3$ as a concrete example, the results presented in this section remain valid for much smaller values of $p$. We leverage the ability of STM to detect the local DSSF directly at or near a single defect (i.e., a vacancy in our simulation). In the chiral KSL, the defect induces a resonant state, observable in the local DSSF as a peak with intrinsic broadening. In contrast, for the non-chiral KSL, the defect generates a truly localized state, which manifests as a sharp peak in the local DSSF, with a width determined entirely by the experimental energy resolution. As we will demonstrate below, the field dependence of the peak width provides a clear distinction between the chiral and the non-chiral KSLs.

\begin{figure*}
    \centering
    \includegraphics[width=0.9\textwidth]{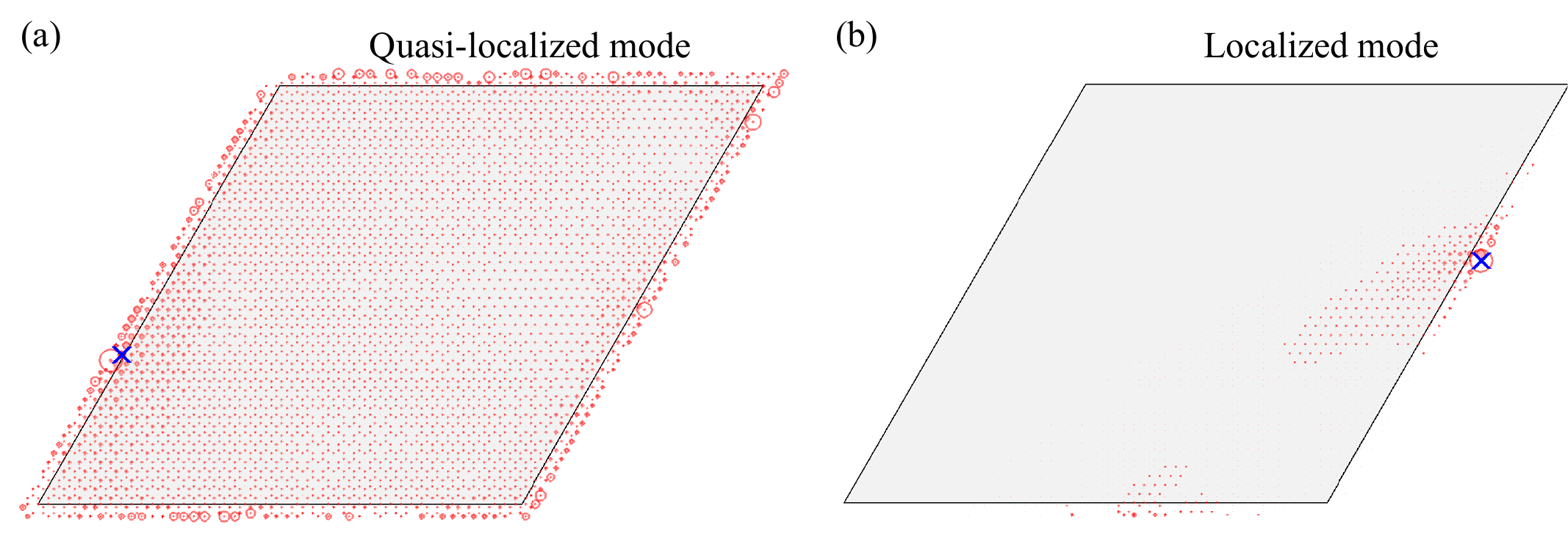}
    \caption{(a) Quasi-localized mode in the chiral case. (b) Localized mode in the non-chiral case. Both cases correspond to the $\omega\simeq 0.05$ peak at $h=0.05$ in the respective DSSFs shown in Figs.~\ref{fig:STM_chiral_disorder}(b) and \ref{fig:STM_chiral_disorder}(d). The blue cross marks the location of the STM tip. The size of each red circle is proportional to the amplitude of the wavefunction ${\bm \Phi}_{s}$ at the given lattice site.}
    \label{fig:wavefunc}
\end{figure*}




\subsection{Chiral Kitaev spin liquid}

We first consider the chiral KSL, and start the investigation by examining the Majorana-fermion spectral functions. In the absence of hybridization, the dangling fermions are localized zero-energy modes, manifesting as sharp $\delta$-peaks at $\omega=0$ in the local spectral function ${\cal A}_{b_j,b_j}(\omega)$, with a total spectral weight of 1. If hybridization is then introduced via Eq.~\eqref{eq:Hz1}, the spectral weight of the dangling fermions spreads across the energy range of the chiral mode, as demonstrated for the clean boundary. The resulting spectrum includes a peak near zero energy---a remnant of the zero-energy peak for bare dangling fermions---and a broad continuum extending up to the top of the boundary band dispersion.

In the presence of disorder, ${\cal A}_{b_j,b_j}(\omega)$ develops a series of peaks superimposed on a continuum background, as shown in Fig.~\ref{fig:STM_chiral_disorder}(a) for disorder strength $p=0.3$. Importantly, a persistent peak always appears around zero energy, originating from the quasi-flat mode at $\rvert k \rvert < k_0$ of the clean boundary. In contrast, the finite-energy peaks appear at different energies depending on the location along the boundary, and correspond to resonances with intrinsic broadening due to the linear-in-$h$ hybridization with the chiral boundary mode. Indeed, by fitting these peaks using a Gaussian function, we find that the peak width increases linearly with $h$, as shown in the inset of Fig.~\ref{fig:STM_chiral_disorder}(a). The real-space wave functions of the corresponding resonance modes are peaked at the location where the given Majorana fermion is excited but extend across the entire lattice [see Fig.~\ref{fig:wavefunc}(a)]. It is important to distinguish the broadening discussed above from the one caused by coupling with the bulk modes. The latter occurs above the bulk gap indicated by the black dashed line in Fig.~\ref{fig:STM_chiral_disorder}(a).

Next, we discuss the local DSSF at a boundary site, which is the convolution of two Majorana-fermion spectral functions [see Eq.~\eqref{eq:local}]. As shown in Fig.~\ref{fig:STM_chiral_disorder}(b), the main features of the local DSSF can still be captured by the convolution involving the narrow $\delta$-peak in ${\cal A}_{b_j,b_j}(\omega)$ around zero energy, similar to the first term in Eq.~\eqref{eq:Sw} for a clean boundary. Therefore, the local DSSF is approximately proportional to ${\cal A}_{c_j,c_j}(\omega)$, whose structure is similar to that of ${\cal A}_{b_j,b_j}(\omega)$. This particular convolution gives rise to peaks around the same positions as in the Majorana-fermion spectral function. Notably, the peak width in the local DSSF is well above the energy resolution used in the calculation and increases linearly with $h$. Note that there are additional peaks originating from the convolution of two finite-energy resonance modes, which could be above the bulk gap. For these peaks, however, it is hard to distinguish the broadening due to the chiral boundary mode and that due to the bulk modes since the dangling fermions couple to both of them through the same Zeeman term. Therefore, for the purpose of detecting the chiral boundary mode, one must focus on the structure of the DSSF below the bulk gap.

It is also worth mentioning that a finite-energy resonance mode already appears around a $60^\circ$ corner of the lattice, i.e., the intersection of two zigzag boundaries [see Fig.~\ref{fig:wavefunc}(a)] in the absence of disorder. The convolution of the zero-energy peak in ${\cal A}_{b_j,b_j}(\omega)$ with this ``corner mode'' leads to an extra peak in the local DSSF.

\begin{figure}[t!]
    \centering
    \includegraphics[width=0.48\textwidth]{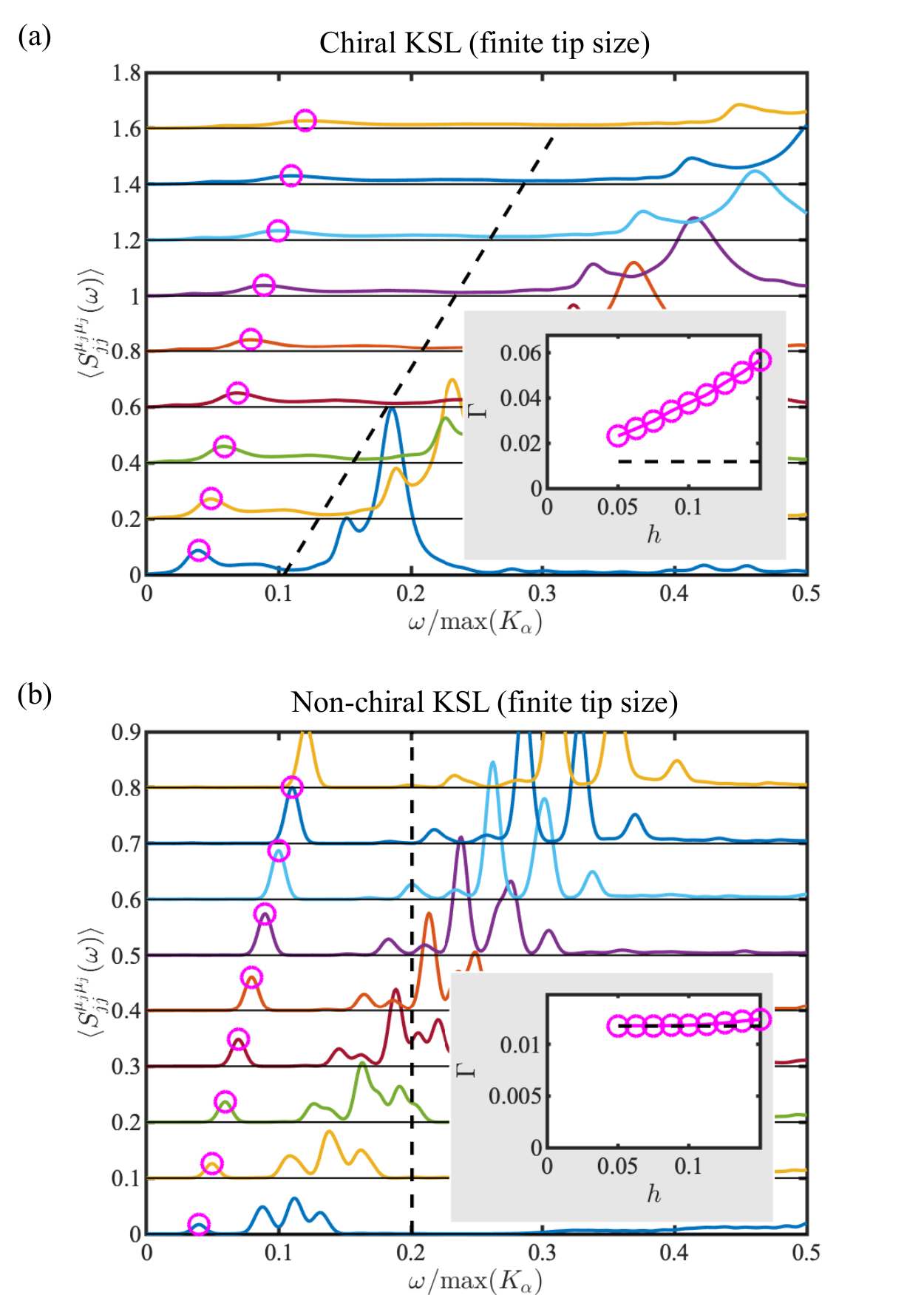}
    \caption{Local DSSF averaged over lattice sites within the effective STM tip size, which is assumed to be five honeycomb lattice constants, for (a) the chiral KSL and (b) the non-chiral KSL. Model parameters and disorder realizations
    are the same as in Fig.~\ref{fig:STM_chiral_disorder}. The black dashed line marks the bulk gap. Inset: HWHM for the peak marked by pink circles in the main panel, with the black dashed line indicating the energy resolution used in the calculation.}
    \label{fig:tip_average}
\end{figure}

The calculation above assumes that STM has atomic-scale spatial resolution. In a real experiment, however, the effective tip resolution spans a few lattice sites. To account for a finite tip size, we average the local DSSF over five adjacent boundary sites. As shown in Fig.~\ref{fig:tip_average}(a), we find that the resonance peaks remain resolved for disorder strength $p = 0.3$. Moreover, both the peak position and the peak width continue to exhibit linear scaling with the field $h$. We finally note that, by averaging over progressively more boundary sites, one eventually crosses over into the disorder-averaged regime discussed in the companion paper~\cite{comp}, where the resonance peaks corresponding to different boundary positions coalesce into a single peak that continues to exhibit linear energy scaling.


\subsection{Non-chiral Kitaev spin liquid}

To contrast with the chiral KSL, we next consider the non-chiral KSL. Following the same approach, we first examine the Majorana-fermion spectral functions. Similar to the chiral case, ${\cal A}_{b_j,b_j}(\omega)$  displays several peaks at discrete energy levels, as shown in  Fig.~\ref{fig:STM_chiral_disorder}(c). Unlike in the chiral case, however, these peaks do not exhibit any intrinsic broadening. Instead, the Gaussian fit indicates that the peak width is determined solely by the energy resolution used in the calculation, meaning that  the peak width remains constant as a function of $h$, as long as the energy is below the bulk gap. We also observe that the wave functions of the corresponding modes are localized in real space, as illustrated in  Fig.~\ref{fig:wavefunc}(b).

Figure~\ref{fig:STM_chiral_disorder}(c) corresponds to the relative boundary orientation shown in Fig.~\ref{fig:local_Sw}(c), in which case there is a gapless boundary mode in the clean boundary limit. Upon introducing disorder, this gapless boundary mode becomes localized, giving rise to peaks in ${\cal A}_{b_j,b_j}(\omega)$ near zero energy. In contrast, for the alternative boundary considered  in Fig.~\ref{fig:local_Sw}(b), where the boundary modes are gapped in the clean boundary limit, the localized modes only appear at finite energy. However, this distinction does not produce observable differences in the local DSSF.

The convolution of these localized boundary modes in the Majorana-fermion spectral function results in a series of discrete $\delta$-peaks in the local DSSF. The energies of these peaks are found by summing the energies of any two distinct $\delta$-peaks in the Majorana-fermion spectral function, in accordance with the Pauli exclusion principle. Therefore, the local DSSF shown in Fig.~\ref{fig:STM_chiral_disorder}(d) does not exhibit a peak around zero energy.

As the field strength increases, the energies of these modes rise linearly, while their peak widths remain constant, determined solely by the energy resolution $\eta = 0.005{\rm max}(K_\alpha)$ (i.e., the imaginary part of the complex frequency $\omega + i \eta$) used in the calculation. This behavior in the non-chiral case is clearly distinct from the chiral case, where the peak width increases linearly with the field strength. Moreover, we confirm that this distinction remains observable after averaging over multiple lattice sites within the effective tip size [see Fig.~\ref{fig:tip_average}(b)]. Finally, the ``corner mode'' mentioned earlier becomes a truly localized mode for the non-chiral KSL, which implies that its convolution with any other mode localized due to disorder results in an additional $\delta$-peak in the local DSSF. 


\section{Discussion}

Diagnosing quantum spin liquids in real materials is a generally difficult endeavor. For chiral spin liquids, one can exploit the existence of a zero-energy chiral boundary mode, which in principle could be detected through anomalous thermal transport measurements. However, multiple attempts to apply this strategy to $\alpha$-RuCl$_3$ have shown that disentangling the magnetic and lattice contributions to the thermal Hall conductivity is highly challenging~\cite{Kasahara18,Yamashita20,Tokoi21,Lefran22,Kasahara22,Bruin22,Czajka23,Zhang23,Kumpei24}.

Therefore, it is natural to look for alternative experimental probes of the chiral boundary mode. Recent advances in STM enable real-space spin spectroscopy, allowing experimentalists to directly observe spin excitations at the boundaries of 2D materials. However, as we discuss in this work, the boundary may also host other low-energy excitations, which can occur even in non-chiral spin liquids. Thus, it is crucial to distinguish between cases where only these boundary modes are present and those where they coexist with a chiral boundary mode.

In the specific case of Kitaev spin liquids, the additional low-energy modes arise from dangling Majorana fermions, which hybridize with the chiral boundary mode whenever the KSL is chiral. The presence of a topologically protected chiral boundary mode can then be detected by exploiting its resistance to localization in the presence of boundary disorder. This topological obstruction to Anderson localization is transferred to the additional boundary modes once they hybridize with the chiral boundary mode. Specifically, the disorder-induced localized boundary modes, which exist already in the absence of the topologically protected chiral mode, become resonances when the chiral mode is present. Since the width of each resonance is proportional to the hybridization, which in turn is linear in the applied magnetic field, the presence of the chiral mode can be detected by verifying that both the positions and the widths of the low-energy resonances scale linearly with the external magnetic field. 

We conclude that STM is one of the most promising experimental probes for identifying chiral spin liquids. Since its effective spatial resolution allows for probing magnetic atoms on scales smaller than 10 nm---with potential further improvements through enhanced tip quality and optimized experimental conditions---we expect this technique to play a crucial role in diagnosing quantum spin liquids in real materials.

\section{Data Availability}
The datasets generated during and/or analyzed during the current study are available from the corresponding author upon reasonable request.

\section{Code Availability}
The codes used during the current study are available from the corresponding author upon reasonable request.

\section*{Acknowledgments}
SSZ thanks Ruixing Zhang for enlightening discussions. This material is based upon work supported by the U.S. Department of Energy, Office of Science, National Quantum Information Science Research Centers, Quantum Science Center.

\section{Author Contributions}

\bibliography{ref}

\end{document}